\documentclass[preprint,showpacs,showkeys,nofootinbib,preprintnumbers,floatfix]{revtex4}

\usepackage{amsmath,epsfig}

\begin{document}

\preprint{hep-th/0309050}

\title{\Large A comment on the self-tuning of cosmological constant 
with deficit angle on a sphere}

\vskip 3cm

\author{Hyun Min Lee}
 \email{minlee@th.physik.uni-bonn.de}
\affiliation{Physikalisches Institut der
Universit\"at Bonn, Nussallee 12, D53115, Bonn, Germany}

\vskip 5cm
\begin{abstract}
In the 6D brane world model with a 4-form flux on a sphere $S^2$ 
for self-tuning the cosmological constant, 
we comment on the fine-tuning problem  
in view of the quantization 
of the dual 2-form flux and the orbifolding case $S^2/Z_2$. 
\end{abstract}


\maketitle

For recent few years, the extra dimensional models  
where the Standard Model(SM) fields reside 
in the hypersurface(so called the brane) of a higher dimensional spacetime 
have drawn much attention, in particular, with the hope of 
solving the cosmological constant problem. 
In these brane world models, there is a
possibility that the SM quantum corrections, which contributes only to 
a brane tension, could be under control by some bulk relaxation mechanism.
We are interested in the {\it self-tuning} solution in the sense 
that a 4D flat solution is always obtained by choosing an integration
constant of the bulk solution without a fine-tuning  
between Lagrangian parameters\cite{ah}.  

In 5D models, it has been shown that the attempt with 
a bulk massless scalar has a hidden fine-tuning in curing the naked 
singularity of the warp factor\cite{nilles} 
and some self-tuning solutions need a particular type 
of the bulk action\cite{kkl} or higher curvature terms\cite{marek}. 
In any case, in 5D self-tuning models, 
changing a brane tension needs a change of the bulk solution. 

On the other hand, in a 6D model with extra dimensions compactified 
on a factorizable $S^2$ with a deficit angle, it has been known that 
changing a brane tension needs only changing a deficit angle  
once there is a bulk tuning through 
the flux\cite{carroll,nav1,cline,quevedo,nav2}. 
In this case, the flux is also responsible for the stabilization of extra 
dimensions. For a concrete example,  
the bulk 2-form flux has been considered\cite{carroll,cline,quevedo,nav2} 
but it has been shown that the quantization condition makes the flux 
dependent on the deficit angle and thus a fine-tuning between bulk and brane 
cosmological constants is indispensible\cite{quevedo,nav2}. 
For other example, however,
when the bulk 4-form flux is considered instead of the 2-form flux, it has
been claimed that there is no flux quantization or no deficit angle dependence
of the 4-form flux\cite{nav2}. Recently, it has been shown that 
this 6D self-tuning model with 2-form or 4-form flux is not different from
the old 4D one with a {\it non-dynamical} 4-form flux from the 4D effective 
theory point of view\cite{nilles2}. 

In this paper, we consider a self-tuning solution with a 3-form field 
$A_{MNP}$($M,N,P=0,1,2,3,5,6$) 
in 6D with extra dimensions compactified on a two-sphere $S^2$ 
with a deficit angle\cite{nav2}. We show explicitly 
that there appears a quantization
condition with the deficit angle dependence 
when the 3-form field couples to a magnetic source. 
We comment more on the fine-tuning problem in this model.

On any six-dimensional manifold, the dual transformation
of the field strength $H_{MNPQ}=\partial_{[M}A_{NPQ]}$ into a 2-form 
field strength $F_{MN}=\partial_{[M}A_{N]}$ is defined as\cite{nakahara} 
\begin{eqnarray}
F_{MN}=\frac{1}{4!}\sqrt{-g_6}\,\epsilon_{MNPQRS}H^{PQRS} \label{dual}
\end{eqnarray} 
where $g_6$ is the 6D metric determinant and $\epsilon_{MNPQRS}$ is 
the Levi-Civita symbol with $\epsilon_{012356}=-\epsilon^{012356}=1$. 
Then, the 6D action with a 3-form field and its coupling 
to both electric and magnetic sources is
\begin{eqnarray}
S&=&\int d^4 x d^2 y\sqrt{-g_6}\bigg(\frac{M^4}{2}R-\Lambda_b
-\sum_{i=1}^2\frac{\sqrt{-g^{(i)}_{4}}}{\sqrt{-g_6}}
\Lambda_i\delta^2(y-y_i)\bigg)
\nonumber \\
&-&\frac{1}{2\cdot 4!}\int H\wedge *H-e\int_{W_3} A_3-g\int_{W_1} A_1
\end{eqnarray} 
where $H=dA_3$, and $*H$ is the Hodge dual of $H$, 
and $A_1$ comes from the definition of the dual field 
strength $F=dA_1$, and $W_3,W_1$ denotes the world volumes of electric
and magnetic sources, respectively. 
Here, $M$ is the 6D fundamental scale, 
$g^{(i)}_4$ are the 4D metric determinants, 
$\Lambda_b,\Lambda_i$ are bulk and 3-brane cosmological
constants and $e,g$ are electric and magnetic charges of sources 
for the 3-form field.

Now let us take the metric ansatz as the direct product of 4D space
and a two-sphere with a deficit angle $2\pi(1-\beta)$, 
\begin{eqnarray}
ds^2=g_{\mu\nu}(x)dx^\mu dx^\nu+\gamma_{mn}(y)dy^m dy^n \label{metric}
\end{eqnarray}
with 
\begin{eqnarray}
\gamma_{mn}(y)dy^m dy^n=R^2_0(d\theta^2+\beta^2 {\rm sin}^2\theta d\phi^2),
\label{extra} 
\end{eqnarray}
and the ansatz for the field strength $H$ as
\begin{eqnarray}
H_{\mu\nu\rho\sigma}=\sqrt{-g}\,\epsilon_{\mu\nu\rho\sigma}E, \ \ \
{\rm others}=0, \label{4form}
\end{eqnarray}
where $E$ is an arbitrary constant. 
Then, the ansatz for $H$ satisfies both the field equation 
and the Bianchi identity for $H$
\begin{eqnarray} 
\partial_M (\sqrt{-g_6}H^{MNPQ})=0, \ \ \ \partial_{[M}H_{NPQR]}=0.
\end{eqnarray}
Moreover, the Einstein equation to be also satisfied is
\begin{eqnarray}
G_{MN}\equiv R_{MN}-\frac{1}{2}Rg_{MN}=\frac{1}{M^4}T_{MN}
\end{eqnarray}
where
\begin{eqnarray}
T_{MN}&=&-\left(\begin{array}{ll}
(\Lambda_b+\frac{1}{2}E^2)g_{\mu\nu} & 0 \\
0 & (\Lambda_b-\frac{1}{2}E^2)\gamma_{mn} \end{array}\right) \nonumber \\
&-&\sum_{i=1}^2\frac{\Lambda_i}{\sqrt{\gamma}}
\left(\begin{array}{ll}
g_{\mu\nu} & 0 \\
0 & 0 \end{array}\right)\delta^2(y-y_i).
\end{eqnarray}
Here, the non-vanishing components of the Einstein tensor\cite{carroll} 
are given by
\begin{eqnarray}
G_{\mu\nu}&=&(R_4)_{\mu\nu}-\frac{1}{2}(R_4+R_2)g_{\mu\nu}, \\
G_{mn}&=&(R_2)_{mn}-\frac{1}{2}(R_4+R_2)\gamma_{mn}
\end{eqnarray}
where $R_4((R_4)_{\mu\nu}), R_2((R_2)_{mn})$ are the Ricci scalars(tensors) 
for the 4D space and the two-sphere, respectively.

Then, for a 4D flat solution with $(R_4)_{\mu\nu}=0$ and $R_4=0$,
the bulk equation gives two conditions  
\begin{eqnarray}
E^2&=&2\Lambda_b, \label{bulk1}\\ 
R^{-2}_0&=&M^{-4}\bigg(\Lambda_b+\frac{1}{2}E^2\bigg), \label{bulk2}
\end{eqnarray}
while the boundary conditions at the branes determine the deficit angle
in terms of brane tensions 
\begin{eqnarray}
2\pi(1-\beta)=\Lambda_1=\Lambda_2.\label{brane}
\end{eqnarray}

At first sight, it seems that there always exists a flat solution for arbitrary 
brane tensions once the 3-form flux takes a value 
satisfying eq.~(\ref{bulk1}). 
Since the field strength of the 3-form takes a value only along the 4D space,
it looks independent of the geometry of extra dimensions such as the deficit 
angle.
Moreover, it was argued that
the fine-tuning between two brane tensions in eq.~(\ref{brane}) 
is avoidable by considering $S^2/Z_2$ instead of $S^2$\cite{nav1,nav2}.
As in the case at hand, since there is no warp factor in the 4D metric part, 
it seems that one does not need to introduce a 4-brane in the action 
even after $Z_2$ orbifolding.   
  
However, this is just the result of disregarding the coupling 
of the 3-form field to the magnetic sources. By duality of eq.~(\ref{dual}), 
the ansatz for the 4-form field strength 
of eq.~(\ref{4form}) becomes the ansatz for the 2-form field strength 
in the spherical coordinate of extra dimensions
\begin{eqnarray}
F_{\theta\phi}=-\sqrt{\gamma}\,\epsilon_{\theta\phi}E, 
\ \ \ {\rm others}=0.
\end{eqnarray}
Then, with the metric (\ref{metric}), 
the 1-form gauge field solution is obtained 
for the  upper and lower hemispheres as
\begin{eqnarray}
A_{\phi,\pm}=\beta ER^2_0(\cos\theta\mp c)
\end{eqnarray}
where $c=1$ from the identity, $\int_{S^2} F_2=\int_{S^1}(A_{1,+}-A_{1,-})$.
Then, the solutions of the gauge field 
must be related by a gauge transformation\cite{quevedo,nav2}:
\begin{eqnarray}
A_{\phi,+}=A_{\phi,-}+\partial_\phi\alpha(\phi)
\end{eqnarray} 
where $\alpha(\phi)=-2\beta ER^2_0\phi$.
Consequently, from the single-valuedness 
of the gauge transformation $e^{ig\alpha(\phi)}$, we get the quantization
condition
\begin{eqnarray}
E=\frac{n}{2g\beta R^2_0} \label{quantum}
\end{eqnarray}
with $n$ an integer.
Then, using this quantization condition and eq.~(\ref{bulk2}), 
the bulk fine-tuning condition 
(\ref{bulk1}) becomes 
\begin{eqnarray}
\frac{n^2}{2g^2\beta^2}=\frac{M^8}{\Lambda_b}.
\end{eqnarray}
Thus, we find that from the quantization of the dual field,
the brane tension also enters the fine-tuning condition via the deficit angle. 

Now let us remark on the possibility with $S^2/Z_2$.
In this case, when we consider the covariant derivative 
for a charged field under the 1-form gauge field, 
the gauge field transforms under the $Z_2$ reflection, 
$\theta\rightarrow\pi-\theta$ and $\phi\rightarrow\phi$, as 
\begin{eqnarray}
A_\theta\rightarrow -A_\theta, \ \ \ A_\phi\rightarrow A_\phi.\label{az2}
\end{eqnarray} 
Thus, the field strength also transforms under $Z_2$ as
\begin{eqnarray} 
F_{\theta\phi}\rightarrow -F_{\theta\phi}.\label{fz2}
\end{eqnarray}
Then, we get the field equation for $F$ as
\begin{eqnarray}
\partial_\theta(\sqrt{\gamma}F^{\theta\phi})
=E\delta(\theta-\frac{\pi}{2})\neq 0. 
\end{eqnarray}
This means that in order to match the boundary condition for $F$, 
we must introduce around the equator 
an extended(4-brane) {\it electric} source under $A_1$, 
which is an extended {\it magnetic} source under $A_3$.  
Then, the 6D action for the dual 2-form becomes  
\begin{eqnarray}
\int d^4x d^2 y\sqrt{-g_6}\bigg(-\frac{1}{4}F_{MN}F^{MN}
-\frac{\sqrt{-g_5}}{\sqrt{-g_6}}A_a J^a\delta(\theta-\frac{\pi}{2})\bigg)
\end{eqnarray}
where $a$ runs over $0,1,2,3,6$, 
and $g_5$ is the determinant of the induced 5D metric on the equator 
with $ds^2_5=ds^2_4+R^2_0\beta^2d\phi^2$, and $J^a=Q\delta^a_\phi$
with an electric charge $Q$ under $A_1$.
Therefore, 
the modified field equation for $A_1$ determines the dual 2-form flux 
in terms of the charge $Q$ as 
\begin{eqnarray}
E=R_0\beta Q. 
\end{eqnarray}
Using eqs.~(\ref{bulk1}) and (\ref{bulk2}),
this result leads to a necessary condition for the charge as 
\begin{eqnarray}
Q=\pm \frac{E}{R_0\beta}=\pm\frac{2\Lambda_b}{\beta M^2}.
\end{eqnarray}
Due to the $Z_2$ property of $F_{\theta\phi}$ (\ref{fz2}), 
the general solution for $A_\phi$ for the upper and lower hemispheres 
is given by
\begin{eqnarray}
A_{\phi,\pm}=\pm\beta ER^2_0(\cos\theta+c_\pm)\label{monopole2}
\end{eqnarray}
where $c_\pm$ are integration constants with $c_+=-c_-$ 
for the $Z_2$ even $A_\phi$.
The 4-brane source term generically contributes to the energy-momentum tensor
as
\begin{eqnarray}
T_{MN}=\bigg[\frac{1}{2}(A_aJ_b+A_bJ_a)-g_{ab}A_c J^c\bigg]\times
\frac{\sqrt{-g_5}}{\sqrt{-g_6}}\delta^a_M\delta^b_N\delta(\theta-\frac{\pi}{2})
\end{eqnarray}
which becomes under our solution 
\begin{eqnarray}
T^0_0=T^i_i=-\frac{Q}{R_0}A_\phi\delta(\theta-\frac{\pi}{2}), \label{em}
\ \ \ {\rm others}=0.
\end{eqnarray}
In order for the charged 4-brane not to contribute to the energy-momentum 
tensor, we need to have $c_+=0$ on the equator 
as the boundary condition for the gauge field. 
However, the Stokes theorem does not hold around the 3-brane,
i.e. $\int_\Sigma F_2\neq\int_{\partial\Sigma} A_1$ where $\Sigma$ is
the infinitesimal surface surrounding the 3-brane.
Then, there are two probable solutions for this: 
one is to modify the field strength $F$, 
and the other is to choose a difference gauge choice with $c_+=-1$ 
at the 3-brane\footnote{The author thanks G. Tasinato for valuable discussion on this issue.}. 

In the former case, the solution for $F_{\theta\phi}$ is supposed
 to be modified with a singular part,
\begin{eqnarray} 
F^s_{\theta\phi}=2\pi\beta E R^2_0 \epsilon_{\theta\phi}\delta^2(y).
\end{eqnarray} 
Then, the field equation for $F$ implies that a charge on the 3-brane 
might be added with the 4D action\cite{gian} 
\begin{eqnarray}
-q\int d^4 x\sqrt{-g_4}F_{\theta\phi}\epsilon^{\theta\phi}
\end{eqnarray}
with $q=-2\pi\beta E R^2_0$. 
However, in this case, it would be indispensible to have the original
solution modified with a $\delta^2(0)$ term
coming from the singular part of $F_{\theta\phi}$ in the energy-momentum tensor.

On the other hand, in the latter case, where there are different 
gauge choices with $c_+=-1$
and $c_+=0$, we can regard the gauge fields to be related to each other 
by a gauge transformation. 
Therefore, the dual 2-form flux on $S^2/Z_2$ is quantized as
\begin{eqnarray}
E_{S^2/Z_2}=\frac{n}{g\beta R^2_0}=2E_{S^2}\label{quantum2}
\end{eqnarray} 
where $n$ is an integer and the subscript of $E$ denotes the case we consider. 
Since the fundamental region on $S^2/Z_2$ is reduced to one hemisphere, 
it is reasonable to
have the magnitude of the flux doubled, compared with the $S^2$ case.
In this case, even if the original metric solution is maintained, 
there appears again a fine-tuning condition between brane and bulk 
cosmological constants in view of the flux quantization (\ref{quantum2}) 
as in the $S^2$ case.

We considered the 6D model with a 4-form flux where the self-tuning
idea may be realized via the deficit angle on $S^2$. 
In this case, we have shown that the coupling of the 3-form field to magnetic
sources is important for the self-tuning issue.
First we have found that there
appears a fine-tuning condition via the quantization of the dual 2-form flux.
We also commented on the case with $S^2/Z_2$ 
for avoiding a fine-tuning between 3-branes.
In this case, 
we showed that a 4-brane charge must be added on the equator 
due to the $Z_2$ property of the gauge field and the dual 2-form flux is also
quantized with a deficit angle dependence 
for maintaining the original metric solution.

\begin{acknowledgments}
I would like to thank S. F\"orste, H. P. Nilles, A. Papazoglou and G. Tasinato 
for helpful discussions. I am supported by the
European Community's Human Potential Programme under contracts
HPRN-CT-2000-00131 Quantum Spacetime, HPRN-CT-2000-00148 Physics Across the
Present Energy Frontier and HPRN-CT-2000-00152 Supersymmetry and the Early
Universe. I am also supported by priority grant 1096 of the Deutsche
Forschungsgemeinschaft.
\end{acknowledgments}

\end{document}